\documentstyle[psfig,conf_iap,]{article}
\begin{document}
\heading{%
%Begin Heading
%
Molecular Gas in the Galactic Halo and Beyond\\
%
%End Heading
} 
\par\medskip\noindent
\author{%
%Begin Author names
Philipp Richter$^{1}$
%End Author names
}
\address{%
%First address
Department of Astronomy, University of Wisconsin/Madison, 475 N. Charter St., Madison, WI 53706, USA
}

\begin{abstract}
I review recent
observations of molecular gas in the
halo of the Milky Way and in the Magellanic Clouds.
Far-ultraviolet absorption line studies of molecular hydrogen (H$_2$)
with ORFEUS and FUSE have unveiled the presence
of a diffuse molecular hydrogen component in
intermediate- and high-velocity clouds in the
Galactic halo. Although the number of
measurements is still quite
small, the data suggests that diffuse H$_2$ in the halo
exists only in clouds that contain a sufficient amount
of heavy elements and dust, on whose surface the H$_2$ formation
proceeds most efficiently. The implications of these
observations for our understanding of the Milky Way halo
are discussed.
\end{abstract}
\section{Introduction}
Studies of interstellar molecules in the Milky Way and
other galaxies are often limited to
observations of carbon monoxide (CO) and more
complex molecules, that are easily observable by
radio emission lines. The most abundant molecule in
the Universe, H$_2$, is difficult to measure directly
because of its homeopolarity.
Diffuse H$_2$ (log $N<21$) can be studied by FUV absorption
spectroscopy of the H$_2$ electronic transitions in
the Lyman- and Werner
bands toward stars and extragalactic sources, but
those measurements (at low redshifts) require satellites in space.
The short-lived {\it Orbiting and Retrievable Far and Extreme Ultraviolet
Spectrometer} (ORFEUS; \cite{barn}) was the first instrument
that allowed a search for H$_2$ absorption outside the disk
of the Milky Way. Today, the more sensitive {\it Far Ultraviolet Spectroscopic
Explorer} (FUSE; \cite{moos}) provides
a large data base to study H$_2$ in environments that are
different from that found in the local Galactic ISM.

\section{H$_2$ Absorption in the Galactic Halo}

\subsection{{\sc ORFEUS} and {\sc FUSE} measurements}
With ORFEUS and 
FUSE, molecular hydrogen has recently been found in
intermediate- and high-velocity H\,{\sc i} clouds (IVCs and HVCs, respectively; see B.P. Wakker,
this conference) in the Galactic halo at $z>0.5$ kpc, representing a diffuse molecular gas
phase above the Galactic plane that was previously undetected in CO emission
or absorption \cite{wak97} \cite{ake}.

\begin{footnotesize}
\begin{center}
\begin{tabular}{llllll}
\multicolumn{6}{l}{{\bf Table 1.} Detections of H$_2$ absorption in IVCs and HVCs} \\
\hline
Target & Cloud Name & $v_{\rm LSR}$ & log $N$(H$_2$)       & log $f^a$     & Instr. \\
       &            & [km\,s$^{-1}$] & [$N$ in cm$^{-2}$]  &               & + Ref.\\
\hline
\multicolumn{6}{c}{Intermediate-Velocity Clouds}\\
\hline
HD\,93521    & IV Arch           & $-60$ & $14.60 \pm 0.35$           & $-4.8$       & O$^b$, \cite{gring}\\
PG\,1351+640 & IV Arch (IV\,16)  & $-50$ & $16.43 \pm 0.31$\,$^c$     & $-3.3$\,$^c$ & F$^b$ \\
PG\,1259+593 & IV Arch           & $-55$ & $14.10^{+0.21}_{-0.44}$\,$^c$     & $-5.4$\,$^c$ & F \\
PG\,0804+761 & LLIV Arch         & $-55$ & $14.71 \pm 0.30$           & $-4.5$       & F, \cite{rich2001a}\\
PG\,0832+675 & LLIV Arch         & $-50$ & $16.10 \pm 0.32$\,$^c$     & $-3.6$\,$^c$ & F \\
Sk\,-68 82   & IVC toward LMC    & $+60$ & $15.65^{+0.32}_{-0.19}$    & $-2.2$       & O \cite{bluhm}\\
Sk\,-60 80   & IVC toward LMC    & $+60$ & present$^d$                & ...$^d$      & F\\
HD\,100340   & IV Spur           & $-30$ & present$^d$                & ...$^d$      & F\\
Mrk\,509     & Complex gp        & $+60$ & present$^d$                & ...$^d$      & F\\
\hline
\multicolumn{6}{c}{High-Velocity Clouds}\\
\hline
Sk\,-68\,82  & HVC toward LMC         & $+130$  & $15.56^{+0.10}_{-0.06}$   & $-3.2$   & O(+F), \cite{rich99} \\
NGC\,3783    & Mag. Stream (LA) & $+240$  & $16.80 \pm 0.10$          & $-2.9$   & F,  \cite{sem} \\
Fairall 9    & Mag. Stream      & $+190$  & $16.40^{+0.26}_{-0.53} $\,$^c$    & $-3.7$\,$^c$ & F\\
\hline
\end{tabular}
\end{center}
\noindent
$^a$ $f=$[$2N($H$_2)/[N($H\,{\sc i})$+2N($H$_2)$]\\
\noindent
$^b$ O=ORFEUS; F=FUSE\\
\noindent
$^c$ Preliminary results\\
\noindent
$^d$ H$_2$ is detected, but no column density has been derived yet\\
\end{footnotesize}
\\

The first detection was that of Richter et al.\,\cite{rich99},
who found
H$_2$ in a HVC in front of the LMC with ORFEUS.
Other detections of H$_2$ with ORFEUS and FUSE in various IVCs
and HVCs have followed since then \cite{gring}
\cite{sem} \cite{rich2001a}.
FUSE observations of extragalactic background sources, and stars
in the Magellanic Clouds and in the Milky Way halo provide
a large data base to explore the molecular content in IVCs and
HVCs over the next years. So far, only a small portion of the data
has been analyzed yet. Table 1 summarizes the recent H$_2$ detections
with ORFEUS and FUSE in IVCs and HVCs. I have included new
(preliminary) results from on-going investigations of atomic and
molecular abundances in the Milky Way halo. As an example, FUSE absorption
line data for Fairall 9, PG\,1259+593 \& PG\,1351+640 are
presented in Fig.\,1. H$_2$ in the Galactic halo is clearly
detected in the Intermediate Velocity Arch (IV Arch) towards PG\,1351+640 \& PG\,1259+593
at approximately $-50$ km\,s$^{-1}$,
and in the Magellanic Stream towards Fairall 9 at $+190$ km\,s$^{-1}$
(for preliminary column densities see Table 1).
In contrast, no H$_2$ is found in the HVC Complex C toward PG\,1259+593,
although the H\,{\sc i} column density in Complex C is relatively high
in this direction (log $N$(H\,{\sc i})=19.92).

\begin{figure}
\centerline{\vbox{
\psfig{figure=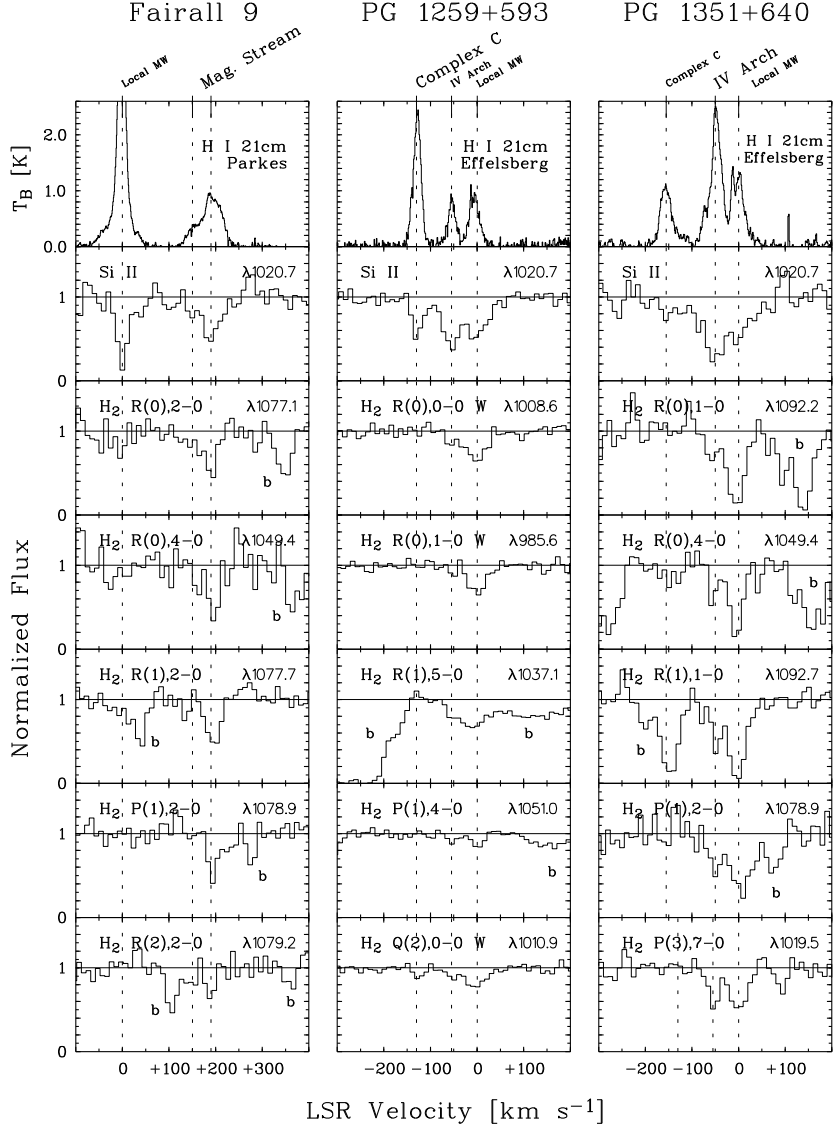,height=16.cm}
}}
\caption[]{\small FUSE data of Fairall\,9, PG\,1259+593 and PG\,1351+640 is presented,
unveiling the presence of H$_2$ absorption in intermediate- and high-velocity gas
in the Milky Way halo. Velocity profiles of
H\,{\sc i} 21cm emission from Effelsberg and Parkes data are plotted 
together with Si\,{\sc ii} and H$_2$ absorption for each individual sight line. 
Toward Fairall 9, H$_2$ absorption at high velocities is present in the Magellanic Stream
at $+190$ km\,s$^{-1}$ (left panel). In contrast, no  H$_2$ absorption is seen in Complex C
toward PG\,1259+593 at $-130$ km\,s$^{-1}$ (middle panel). H$_2$ absorption at intermediate velocities
is seen in the IV Arch toward PG\,1259+593 at $-55$ km\,s$^{-1}$ (middle panel), and toward PG\,1351+640
near $-50$ km\,s$^{-1}$ (right panel). Blending lines from other species are marked with `b'.
}
\end{figure}

\subsection{Molecular Fractions and Metallicity}
The presence of molecular material in Galactic halo clouds is
somewhat suprising, given the expected highly diffuse and low-density nature
of these clouds.
H\,{\sc i} column densities of these clouds are
all below $10^{20}$ cm$^{-2}$, as derived by H\,{\sc i} 21cm data \cite{wak20},
giving no evidence for the existence of dense clumps
along the sight-lines where H$_2$ is found, but beam-smearing effects
must be considered carefully.
H$_2$ column densities vary between $\sim 10^{17}$ and $\sim 10^{14}$ cm$^{-2}$, the
latter roughly being the detection limit for H$_2$ absorption line
studies with FUSE. Therefore, the fraction of hydrogen in molecular
form $f=$[$2N($H$_2)/[N($H\,{\sc i})$+2N($H$_2)$]
is found to be generally low in IVCs and HVCs
(see Table 1).

H$_2$ formation proceeds most efficent on the
surface of dust grains, where the H$_2$ can release part of its binding
energy of 4.5 eV. If dust is not available, H$_2$ can form in the gas phase,
but this process is very slow \cite{black}. The relation between
dust abundance (as measured by the colour excess $E(B-V)$) and
the H$_2$ column density is well established in the Milky Way
\cite{savage}. In low-metallicity environments, such as
the Magellanic Clouds, the fractional abundance of molecular
hydrogen is lower due to the smaller dust-to-gas ratio in the
Clouds \cite{rich2}
\cite{tum} (see also section 3).
Interestingly, the Galactic halo clouds follow a similar trend:
H$_2$ in the Milky Way halo is found mainly in gas that has
nearly solar abundances (i.e., IVCs; see Table 1 and Fig.\,1), except
for the Magellanic Stream, where H$_2$ is found in two of two
observed sightlines at $\sim 0.3$ solar metallicity \cite{sem}.
At the lower end of the metallicity scale in Galactic halo clouds,
no molecular hydrogen is found in Complex C (e.g., toward
PG1259+593; see Fig.\,1) at
$\sim 0.1$ solar metallicity along several sight lines, although the
H\,{\sc i} column densities are in most cases significantly higher
than in IVCs, in which H$_2$ {\it is} detected. The non-detection
of H$_2$ in Complex C therefore implies that this metal-poor
HVC does not contain significant amounts of dust. This is in agreement
with the lack of substantial depletion of Fe and Si in Complex C
\cite{rich2001b}.

\subsection{H$_2$ Fomation and Dissociation in the Milky Way Halo}

Sembach et al.\,\cite{sem} suggested that the molecular hydrogen found in
the Magellanic Stream survived the tidal stripping from the Small
Magellanic Cloud, and thus was brought into the Milky Way halo
from outside. With respect to other halo clouds containing H$_2$,
however, the formation of molecular gas {\it within} the halo
appears to be as likely. If, for example, IVCs represent the cooled
backflow of a Galactic fountain 
\cite{shap}, the H$_2$ found in IVCs must have formed in situ. 

In a steady state, the H$_2$ grain formation and photo-dissociation
are equilibriated (see Spitzer \cite{spitz}, page 125). As preliminary 
calculations show, it is very difficult to describe the
observations of H$_2$ in Galactic halo clouds in an H$_2$
formation-dissociation equlibrium model. 
On the one hand, the relatively low H\,{\sc i} column densities found in
the Galactic halo clouds (log $N<20$) require that the
H$_2$ resides in very small ($<0.1$ pc) clumps that would 
have a relatively low sky-covering factor. On the other
hand, H$_2$ in
IVCs seems to be surprisingly widespread, requiring
at large number of such clumps in the lower Galactic halo.
Another possibilty, however, is that the observed H$_2$ is {\it not} in
formation-dissociation equilibrium, but represents 
molecular gas that has formed in dense, compact clouds,
but then was dispersed into larger volume, e.g., by 
diffusion. If so, the low-density H$_2$ gas is being
photo-dissociated on a time scale of 
$t_{\rm diss.}=(\langle k \rangle \beta_{0,{\rm halo}})^{-1} \approx 10^4$ years.
\footnote{$\langle k \rangle=0.11$ denotes the possibility that an H$_2$ molecule
is dissociated after photo-absorption; we assume that the 
photo-absorption rate in the halo is roughly $1-10$ percent
of that in the Milky Way disk, thus $\beta_{0,{\rm halo}} \approx 0.5-5.0 
\times 10^{-11}$ s$^{-1}$.}
This time-scale is remarkibly short; possibly,
Galactic halo clouds are constantly forming and dispersing
diffuse molecular cloud cores, preventing the formation of CO 
bearing clouds and star formation. If some of those clumps can
survive, however, they might form a population of young
metal-rich halo stars. Interestingly, such stars are indeed 
observed \cite{con}.

\section{H$_2$ in the Magellanic Clouds}
With ORFEUS and FUSE, it is now 
possible to study molecular hydrogen also beyond the Milky Way and
its halo, e.g. in the Magellanic Clouds. Richter \cite{rich2} concluded from
ORFEUS data that the molecular hydrogen fraction in the Clouds is
reduced compared to the Milky Way 
due to the lower dust content and the higher UV radiation field.
This idea is supported by the more extensive and more accurate FUSE data
\cite{tum}. Upcoming FUSE observations of  Magellanic
Cloud stars that have high extinction will help to understand the
transition from the diffuse into the dense molecular gas phase in low
metallicity environments.

\section{Molecular Clumps in the Halo as Dark Matter Candiates}

It has been proposed that extremely dense molecular
'clumpuscules' in the outer Galaxy and/or in the Galactic halo might
serve as reservoir of cold baryonic dark matter in spiral 
galaxies 
\cite{pfenn} .
From the presence of the diffuse extreme $\gamma$-ray ($>100$ MeV)
emission from the Milky Way halo (as measured by EGRET; see
Kalberla et al.\,\cite{kalb}) have
proposed that the observed $\gamma$-ray flux is caused by interaction between
cosmic rays and dense molecular gas clumps by the $\pi_0$-decay.
These clumps could also account for the extreme scattering
events (ESE) that are found toward quasars \cite{fied}.
No {\it direct} observational evidence for such a extremely compact molecular
gas phase in the halo has been found so far, but UV absorption line
studies of extragalactic background sources with FUSE and
other instruments  will be used to search for
evidence of such clumps, which would manifest themselves by
partly or fully absorbing the UV continuum from an extragalactic 
source, if they coincidentally move into the line of sight.

\section{Concluding Remarks}

FUV absorption spectroscopy with FUSE and ORFEUS has unveiled the
presence of a diffuse molecular hydrogen component in the cores of
Galactic halo clouds. Further absorption line studies are necessary
to understand the nature of this gas. H$_2$ absorption spectroscopy
of halo clouds might serve as a powerful diagnostic tool to study
the physical conditions in the Milky Way halo and might help to
obtain new insights on the formation and dissociation of H$_2$
under conditions that are different from those in the Galactic disk.

\acknowledgements{Part of this work is based on data obtained for the
the Guaranteed Time Team by the NASA-CNES-CSA FUSE
mission operated by the Johns Hopkins University.
Financial support has been provided by NASA
contract NAS5-32985. 
I thank B.D. Savage and B.P. Wakker
for helpful comments.
}

\vfill
\end{document}